\documentclass[10pt,letterpaper]{article}
\usepackage[top=0.85in,left=2.75in,footskip=0.75in,marginparwidth=2in]{geometry}

\usepackage[utf8]{inputenc}

\usepackage{cite}

\usepackage{hyperref}
\usepackage{booktabs}

\usepackage{microtype}
\usepackage{longtable}
\DisableLigatures[f]{encoding = *, family = * }

\raggedright
\usepackage{changepage}

\usepackage[aboveskip=1pt,labelfont=bf,labelsep=period,singlelinecheck=off]{caption}

\makeatletter
\renewcommand{\@biblabel}[1]{\quad#1.}
\makeatother

\usepackage{lastpage,fancyhdr,graphicx}
\usepackage{epstopdf}
\fancyheadoffset[L]{2.25in}
\fancyfootoffset[L]{2.25in}

\usepackage{color}

\definecolor{Gray}{gray}{.25}

\usepackage{graphicx}

\usepackage{sidecap}

\usepackage{wrapfig}
\usepackage[pscoord]{eso-pic}
\usepackage[fulladjust]{marginnote}
\reversemarginpar

\begin{document}
\vspace*{0.35in}

\begin{flushleft}
{\Large
\textbf\newline{An Empirically Evaluated Checklist for Surveys in Software Engineering}
}
\newline
\\
Jefferson Seide Moll\'eri\textsuperscript{1},
Kai Petersen\textsuperscript{1},
Emilia Mendes\textsuperscript{1}
\\
\bigskip
\bf{1} BTH - Blekinge Tekniska H\"ogskola
\\
\bigskip
* jefferson.molleri@bth.se

\end{flushleft}

\section*{Abstract}
\textbf{Context:} Over the past decade Software Engineering research has seen a steady increase in survey-based studies, and there are several guidelines providing support for those willing to carry out surveys. The need for auditing survey research has been raised in the literature. Checklists have been used to assess different types of empirical studies, such as experiments and case studies.

\textbf{Objective:} This paper proposes a checklist to support the design and assessment of survey-based research in software engineering grounded in existing guidelines for survey research. We further evaluated the checklist in the research practice context.

\textbf{Method:} To construct the checklist, we systematically aggregated knowledge from 14 methodological papers supporting survey-based research in software engineering. We identified the key stages of the survey process and its recommended practices through thematic analysis and vote counting. To improve our initially designed checklist we evaluated it using a mixed evaluation approach involving experienced researchers.

\textbf{Results:} The evaluation provided insights regarding limitations of the checklist in relation to its understanding and objectivity. In particular, 19 of the 38 checklist items were improved according to the feedback received from its evaluation. Finally, a discussion on how to use the checklist and what its implications are for research practice is also provided.

\textbf{Conclusion:} The proposed checklist is an instrument suitable for auditing survey reports as well as a support tool to guide ongoing research with regard to the survey design process.


\section*{Introduction}
A survey is a widely deployed research method in the area of Software Engineering (SE) and an increase in its usage has been highlighted by, e.g., Punter et al. \cite{punter2003conducting}. Its purpose is to investigate a population, in order to construct explanatory models \cite{babbie1973survey,wohlin2012experimentation} or to validate knowledge \cite{kitchenham2001principles,kitchenham1997desmet}. Survey research is often employed when there is a need to study a large set of variables \cite{wohlin2012experimentation} or to perform a retrospective analysis \cite{pfleeger1995experimental}. It may be used to draw conclusions based on both quantitative and qualitative data \cite{dawson2005projects}.

Researchers have highlighted various challenges during the survey process. Common challenges are the formulation of questions \cite{torchiano2013six}, so to avoid shortcomings (e.g., introducing bias inside questions \cite{wohlin2012experimentation}), and the identification of invalid responses \cite{yang2016survey}. Other challenges are related to the recruitment of participants, such as how to obtain a sufficient number of responses and how to prevent high drop-out rates \cite{akbar2012framework,galster2014exploring}.

The need for improving the completeness of reporting survey-based research, in particular with respect to the definition of the population and sampling strategies is evidenced by Stavru \cite{stavru2014critical}. Furthermore, Stavru pointed out a lack of checklists for auditing surveys in SE which could be of help to both researchers conducting survey research as well as to those evaluating and reviewing the research.

Motivated by these needs, we employed an empirical approach to constructing and evaluating an assessment checklist\footnote{The resulting instrument is further detailed in Appendix \ref{sec:appendixChecklist}.}. Such approach comprises of two steps (see Sections \ref{sec:constChecklist} and \ref{sec:evaluationProfessional}):

First, we detail the process to \textbf{construct a checklist} to assess survey research in SE. The method used to derive the checklist was guided by two principles: (a) identify existing guidelines for survey research; in the context of SE, 14 methodological papers have been considered; (b) elicit the process stages, recommended practices, and related rationales. Those rationales support the cost-effectiveness analysis of employing a set of related practices.

The method for systematically deriving the checklist was based on thematic analysis \cite{cruzes2011recommended}. Vote counting was applied to the themes identified in order to compute the frequency in which they occurred. Further, a co-occurrence was obtained through a relationship matrix relating different categories (e.g., practices versus rationales).

Later, we \textbf{evaluated the checklist} by using a mixed approach \cite{kitchenham1997desmet}. The evaluation process involved two distinct phases: (a) to apply the checklist on a set of published survey reports and register the assessment scores, and (b) to verify the results of this assessment with the corresponding authors of those survey reports.

The assessment produced a compliance coefficient for the selected studies in relation to each of the checklist items. We further investigated the authors' feedback in order to understand patterns we identified in the assessment scores. We also collected and addressed suggestions from the experts to improve the checklist instrument.

The remainder of the paper is structured as follows: Section \ref{sec:background} describes the background and related work. Section \ref{sec:constChecklist} details the systematic approach we used to construct the checklist. The evaluation of the checklist in research practice context is presented in Section \ref{sec:evaluationProfessional}. Section \ref{sec:discussion} discusses the findings and finally, and Section \ref{sec:conclusions} concludes the paper.

\section{Background \& Related Work}
\label{sec:background}

We first present existing survey guidelines that are subject-independent or that have been proposed in other fields. Thereafter, an overview of SE specific guidelines is provided. After giving an overview of the guidelines we describe the literature on survey assessment. Finally, we looked into checklists proposed to support Empirical Software Engineering (ESE).

\subsection{Survey process and guidelines}
\label{sec:surveyguidelinesGeneral}

Survey as a research method has been established in social research for half a century. It has been employed in several academic fields, such as health care, politics, psychology, and sociology \cite{converse2017survey}. As a consequence, methodological knowledge on surveys has been published first in these fields.

As the survey research method matured, cross-field guidelines appeared (e.g., \cite{fowler2013survey,alreck1994survey,fink2003survey}). These publications aimed to provide methodological support independent from the subject of research. Nevertheless, it is not uncommon for their mentioned practices to focus on the social aspects of research.

Those guidelines \cite{fowler2013survey,alreck1994survey,fink2003survey} describe a survey-research process comprising a set of stages, such as question design, sampling, data collection, instrument evaluation, measuring and data analysis \cite{fowler2013survey}. Survey research is acknowledged for being flexible, although the process stages are often conducted sequentially. 

It is worth mentioning that the survey process is a complete research method (i.e., including planning, execution, analysis, and reporting); the survey data collection instrument is called a questionnaire.

In addition to describing the process, the guidelines also recommend best practices based on desirable attributes for high-quality surveys. Such quality is based on evaluation of the produced evidence (e.g., precision, credibility), ethical issues (e.g., consent, privacy) and mitigating validity threats (e.g., sample error, non-responses) \cite{fowler2013survey,lavrakas2008encyclopedia}.

\subsection{Survey guidelines in Software Engineering}
\label{sec:surveyguidelinesSE}

The need for specific and tailored guidelines to conduct empirical research in the context of SE has been pointed out, e.g. \cite{perry2000empirical,sjoberg2007future,wohlin2016there}.
This demand is especially relevant to formal experiments and case studies, due to the popularity of such methods, but also applies to survey-based research. Methodological support for surveys in SE first appeared around the 1990s \cite{molleri2018cerse}.

Three main guidelines \cite{kitchenham2001principles,kasunic2005designing,linaker2015guidelines} detail the survey research process in the SE field. They jointly provide a comprehensive structure for the research process, despite differing slightly from each other. Major differences are in relation to the breakdown structure of process stages and the recommended practices provided.

Besides these three main publications, a series of additional studies extend the guidance to particular stages of the survey process. For example, the challenges of identifying the target audience and establishing a sampling frame are discussed in \cite{de2013would,de2014sampling,de2014towards,de2015investigating}. The recommended practices in this set of papers are complementary, although some partially overlap.

Other studies provide lessons learned from carrying out the process in different contexts:
\begin{itemize}
    \item Punter et al. \cite{punter2003conducting} focus on self-administered online surveys and address issues such as monitoring real-time responses, identifying the reasons for dropouts and encouraging participants to complete a survey instrument;
    \item Ciolkowski et al. \cite{ciolkowski2003practical} addresses practical issues related to the process itself, such as managing resources and ensuring that deviations do not threaten the completion of the entire process; and
    \item Cater et al. \cite{cater2005addressing} address replication challenges, such as updating a survey instrument and comparing the results.
\end{itemize}

Additional references for survey-based research are provided in our previous works \cite{MolleriPM16,molleri2018cerse}.

\subsection{Survey assessment}
\label{sec:evaluations}

When reviewing existing guidelines (see Section \ref{sec:surveyguidelinesSE}) we found out that several researchers highlighted the need for an instrument to audit survey research in SE context. This need is further stressed by the lack of reporting of the employed criteria to assess survey research \cite{stavru2014critical}.

Stavru's work \cite{stavru2014critical} provides a critical review of surveys in the area of agile software development. In order to carry out the review, Stavru used 21 criteria by which the thoroughness in reporting surveys was assessed. These criteria were extracted from different sources, cf. \cite{kitchenham2001principles,kasunic2005designing,ciolkowski2003practical,pinsonneault1993survey,shenton2004strategies,malhotra1998assessment}. Note that the method of eliciting the criteria was not detailed.

Stavru also highlighted that the different criteria were not equally important, and rated them on a scale from one to five. The most important criteria that ought to be documented were:

\begin{itemize}
    \item Sampling frame, method, and size
    \item Response rate
    \item Assessment of a survey's trustworthiness 
    \item Survey process
    \item Conceptual model comprising of the constructs investigated (e.g., variables and their relations)
    \item Target population
    \item Questionnaire design
\end{itemize}

\subsection{Checklists in Software Engineering}
\label{sec:checklists}

Checklists have been proposed for various research methods with a specific focus on their usage in the SE context. Looking at the ways in which checklists were built, researchers most often based the construction of a new checklist upon existing ones (cf. \cite{HostR07,WieringaCDMP12}).

As an example, Kitchenham et al. \cite{KitchenhamSBBDHPR10} combined two checklists \cite{DybaD08,KitchenhamBTNLPB10} to assess experiments and to evaluate whether researchers may use them objectively. Their findings indicate that a larger number of reviewers was needed (eight) to reliably assess studies using their checklist, which could be improved by having researchers conduct reviews in pairs (cf. \cite{KitchenhamSBBDHPR10}). Additional checklists proposed for assessing experiments are, e.g., \cite{wohlin2012experimentation,JedlitschkaP05,KitchenhamSBBDHPR10}.

H\"ost and Runeson \cite{HostR07} put forward a checklist for case study research, divided according to the research stages, including design, preparation for data \& evidence gathering, data analysis, as well as reporting. To ease a reviewer's task, a condensed checklist abstracting the original checklist has been suggested to reduce the number of items to be checked. Additional checklists for case study research in SE are found in, e.g., \cite{kitchenham1995case,perry2004case,wohlin2012experimentation}.

Wieringa \cite{Wieringa12} observed that the individual checklists with the same focus differed, which may result in confusions for reviewers. The author highlights the need to find common checklist items across research types as they may share specific aspects. Thus, Wieringa et al. \cite{WieringaCDMP12} used existing checklists (e.g., \cite{JedlitschkaP05,HostR07,consort2010statement}) for experiments and case studies as a basis to synthesize an unified checklist. Later, the authors evaluated their checklist by having them used by PhD students and researchers in different research groups, as well as by conference participants.

Stavru's \cite{stavru2014critical} filled a gap in the existing body of knowledge by complementing the set of available checklists with a set of criteria for assessing survey research. No other checklists to assess surveys were identified in our systematic literature search (cf. \cite{MolleriPM16}).

Great emphasis was placed upon (a) basing the checklist on existing literature, and (b) following a systematic approach to eliciting checklist items \cite{stavru2014critical,Wieringa12}. Thus, our work complements the above-mentioned by deriving and evaluating an assessment checklist grounded in existing guidelines for survey research.

\section{Step 1. Construction of the checklist}
\label{sec:constChecklist}

The first step of our research approach entailed the systematic construction of the checklist. Three sub-contributions are made that ultimately lead to the checklist proposed:

\begin{enumerate}
    \item[C1] \textit{Consolidation of survey processes and decision points:} We present a consolidated survey process based on existing guidelines. Key decisions points and implications of decision-making are highlighted. For example, a key decision in a survey process is the type of sampling used, which impacts participant recruitment and data analysis. Our checklist has to be adapted depending on the decisions taken.
    \item[C2] \textit{Extraction of recommended practices and their mapping to the survey process:} We extracted the recommended practices to be carried out during a survey research process, which were later mapped to the research process identified in C1. Mapping the practices to the main stages aids researchers in the planning of surveys, as it indicates in which process step a practice is executed and where its impact needs to be considered.
    \item[C3] \textit{Extraction of rationales for the recommended practices:} The reasons for considering existing survey research practices should be motivated by a rationale, thus making the value of adopting such practices explicit. This is particularly pressing because a survey's cost-effectiveness is an important consideration. Thus, understanding the rationales for the recommended practices supports the cost analysis of a practice and its effectiveness (i.e., the rationale regarding the value a given practice adds to the survey research).
\end{enumerate}

\subsection{Method}
\label{sec:constChecklistMethod}

\subsubsection{Research questions}
\label{sec:constChecklistRQ}

We formulated three research questions corresponding respectively to each of the three contributions stated above, as follows:
\begin{enumerate}
    \item[RQ1] Which stages and key decisions are specified for the survey process (C1)?
    \item[RQ2] Which practices are suggested and how do they map to the stages of the research process (C2)?
    \item[RQ3] What is the rationale for conducting the respective recommended practices (C3)?
\end{enumerate}

\subsection{Study identification and selection}
\label{selection}

In order to select an appropriate set of primary studies, we used evidence from our previous studies \cite{molleri2018cerse,MolleriPM16} identifying methodological papers for survey-based research in SE. 

A set of eight papers provided the main guidelines covering all the stages of a survey research process \cite{kitchenham2001principles,kasunic2005designing,linaker2015guidelines}. We assume that these core papers were likely to hold all the information needed to derive our checklist, i.e., recommended practices and reasons for their adoption.

Given that only a few papers covering the complete methodology may not cover recommended practices and reasons for their adoption sufficiently, we completed the core set with six additional additional supporting papers \cite{punter2003conducting,de2014sampling,de2014towards,de2015investigating,ciolkowski2003practical,cater2005addressing} addressing specific stages of survey research such as sampling, instrument design, and validation, recruitment and response management.

\subsubsection{Data extraction and analysis}
\label{analysis}

We employed a systematic process based on thematic synthesis \cite{cruzes2011recommended} to extract data from and to analyze the primary studies. First, we aggregated the papers in a common list using Atlas.ti \cite{friese2012atlas} - a qualitative data analysis software - for text interpretation. Then, we incrementally read the papers, collected text segments and aggregated them into themes. The themes were classified into three major categories:

\begin{enumerate}
    \item Process stages, e.g., data analysis.
    \item Recommended practices, e.g., identify reasons for non-responses.
    \item Rationale attributes, e.g., representativeness.
\end{enumerate}

The terminology for initial codes is derived from the three guidelines analyzed first (i.e., \cite{kitchenham2001principles,kasunic2005designing,linaker2015guidelines}). Later, we iteratively improved and updated the initial coding according to the different views presented by the additional sources. The segments are characterized by a level of granularity of one paragraph, notwithstanding a single paragraph is often associated with more than one theme. Paragraphs containing no relevant information were associated with no theme.

Successive iterations of the process further refined the theme set, by combining or merging synonyms and aggregating related themes into families (e.g., representativeness is part of the external validity rationale). We also removed duplicates and combined successive occurrences of the same theme in larger segments, thus comprising several paragraphs.

\begin{figure}
  \centering
  \includegraphics[width=1\textwidth]{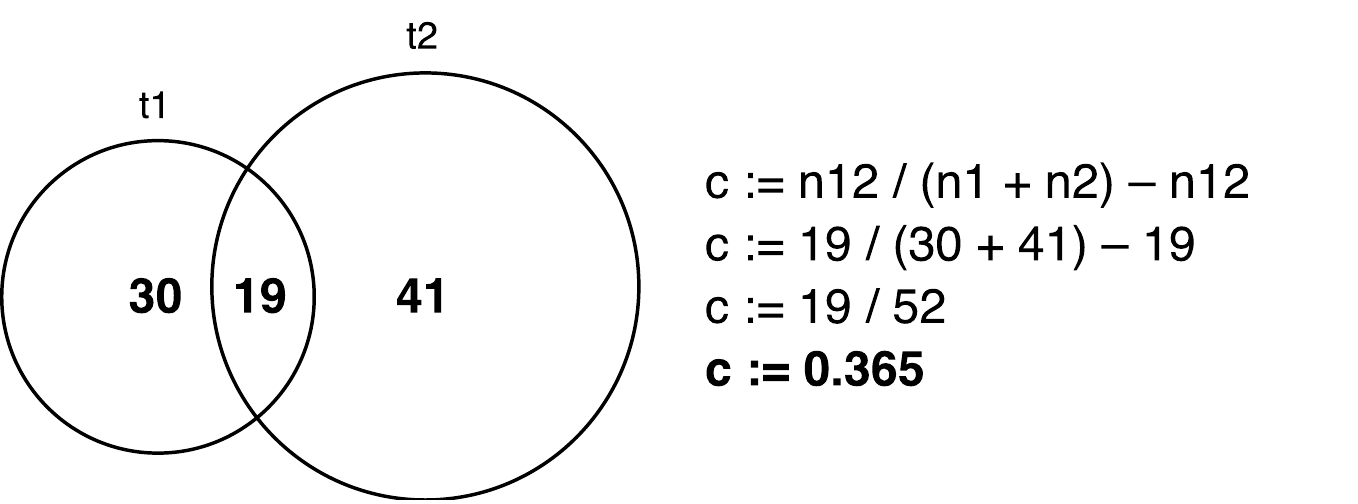}
  \caption{Example for computation of the co-occurrence coefficient, given that t1 occurs 30 times in the data set, t2 occurs 41 times, and they simultaneously occur 19 times.}
  \label{fig:coocurrenceExample}
\end{figure}

Finally, we computed a co-occurrence coefficient \cite{contreras2011examining} to analyze how frequently two related terms occurred alongside each other. This coefficient is calculated as follows: $c := n_{1,2}/(n_1 + n_2) - n_{1,2}$, whereas $n_1$ and $n_2$ are the vote-counting frequencies of two themes $t_1$ and $t_2$ respectively, and $n_{1,2}$ is the joint frequency, i.e., how many times the two themes co-occur. An example of the coefficient computation is given in Figure \ref{fig:coocurrenceExample}.

The resulting relationships are presented in a co-occurrence matrix \cite{friese2012atlas}, where the cells are filled in grey-tones according to the coefficient value (see Section \ref{practices}). Darker cells represent a stronger co-occurrence between two themes. We opted for normalizing the themes in each matrix row since our analysis relates mainly to only comparing themes within the same category. Our normalized coefficient range from 0-100, whereas 100 relates to the maximum occurrence in the corresponding group, and 0 corresponds to no occurrence.

\subsection{Threats to validity}

\textbf{Construct validity.} To ensure a similar understanding and reduce research bias on the thematic analysis, we piloted the coding process between the three authors. The results imply a fair agreement, i.e., in average, 46.5\% of the themes are similar, although worded differently. Further, based on the reflections of the pilot study, the first author coded the remaining papers. The co-occurrence coefficient used in the analysis takes into consideration the position and size of sentences but is prone to non-significant values when comparing themes that differ largely in size \cite{friese2012atlas}. We partially address this potential bias by normalizing the values within the same row.

\textbf{Internal validity.} Internal validity relates to factors affecting the outcome of the study not accounted for by the researchers. One threat is the bias in interpreting the findings. Hence, at each stage of the research, the intermediate results were discussed among the researchers (observer triangulation). 

Our data set consists of 14 papers gathered from a previous literature study \cite{MolleriPM16}. We relied mostly on the original study design to the search and selection stages, using its reported evidence to collect our relevant set. We employed structured reading and coding to analyze the data set, producing themes and higher-level categories. The first author conducted the data extraction and analysis, further discussing the resulting themes with the other co-authors. We trust that this iterative process minimized the judgment bias of a solo researcher.

\textbf{Conclusion validity.} One threat to conclusion validity is whether the data based on which the survey was created is complete. The additional 6 papers in our selection (see Section \ref{selection}) complemented the study results with 10 extra practices and 1 rationale. Those extra themes are related to particular challenges a researcher may face during the process, namely a large sample frame \cite{de2014sampling,de2014towards}, managing online surveys \cite{punter2003conducting}, and survey replication \cite{cater2005addressing}. By adding guidelines very specific to the individual stages of survey research increased confidence in the results, despite being limited by the availability of the literature for each stage. 

\textbf{External validity.} The resulting themes and frequencies were extracted from relevant methodological guidance for SE. However, we cannot assume that the practices and rationales identified are only important for this field. Moreover, there is the possibility of identifying a valid theme outside of our data set, e.g., a non-selected paper or the practical experiences of a researcher. We, therefore, conducted a evaluation of our proposed checklist with SE researchers, thus investigating its appropriateness and identifying potential improvements (see Section \ref{sec:evaluationProfessional}).

\subsection{Results}

The following section is structured according to the three contributions proposed in Section \ref{sec:constChecklistMethod}. Each section is, in turn, broken down into its units of analysis (i.e., decision-making points, process stages, and rationale aggregation).

\subsubsection*{C1. Survey Research Process}
\label{sec:process}

\begin{figure}
  \centering
  \includegraphics[width=1\textwidth]{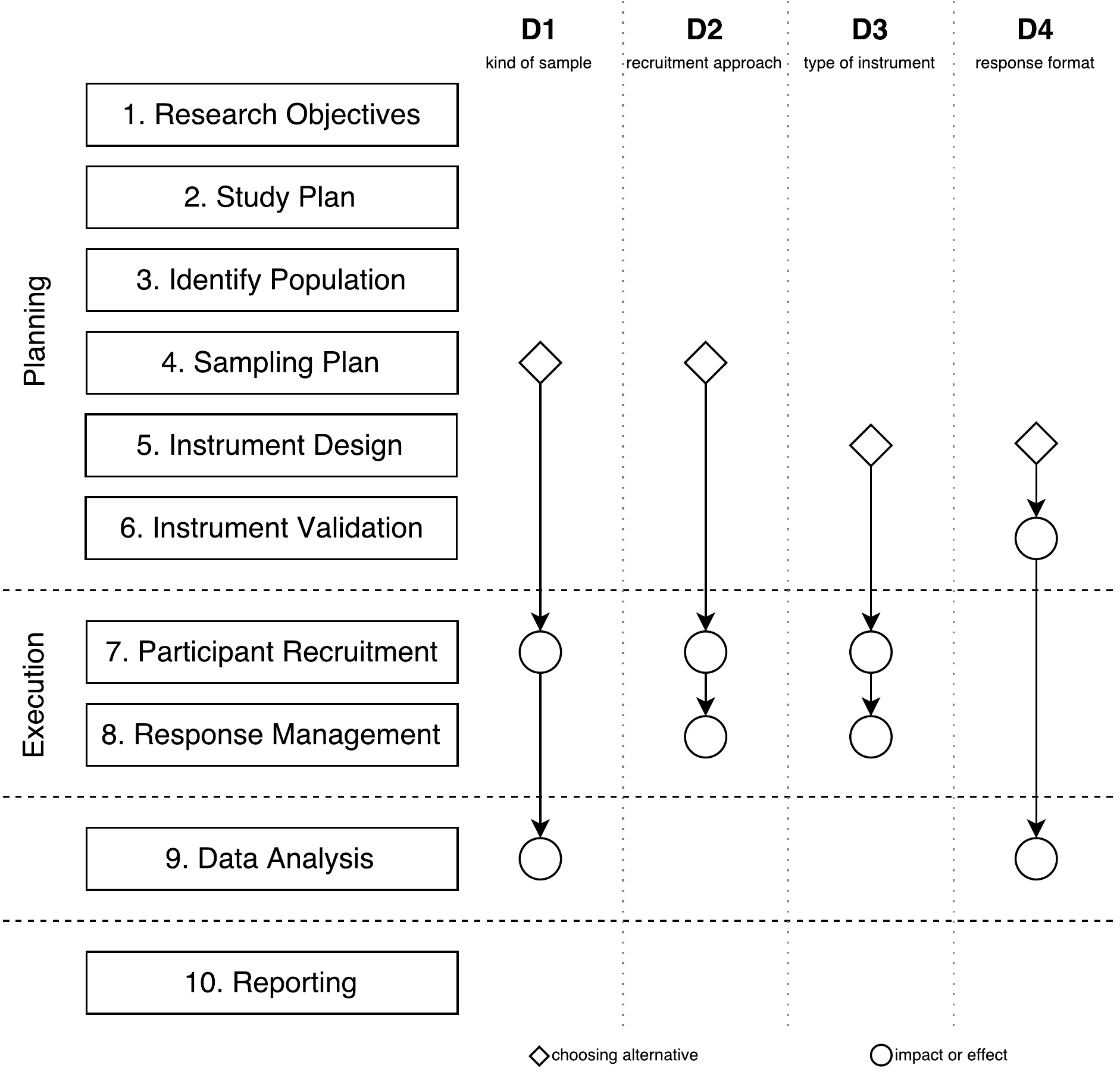}
  \caption{Model for the survey process, together with the decision points that could impact on the research. The diamond symbols ($\diamond$) highlights the stages in which the conditional alternatives should be chosen, and the circles ($\circ$) mark the stages affected by those decisions.}
  \label{fig:surveyProcess}
\end{figure}

Figure \ref{fig:surveyProcess} presents an aggregation of the survey research processes described by Kitchenham \& Pfleeger \cite{kitchenham2001principles}, Kasunic \cite{kasunic2005designing}, and Lin\r{a}ker \emph{et al.} \cite{linaker2015guidelines}. Although the main stages (and terminology) slightly differ among the guidelines, the processes are similar and follow a sequential flow. We adopted the view from \cite{kitchenham2001principles} to describe the execution phase comprising two stages, one for recruiting participants and one for administering responses.

We also identified key decision-making points during the process, two of which should be addressed during the sampling stage (i.e., D1 and D2) and two others during the instrument design stage (i.e., D3 and D4). Those conditional nodes require researchers to make decisions regarding a survey's research design that can potentially impact the subsequent stages.

\begin{enumerate}
    \item[D1] \emph{What kind of sample is selected?}\\Depending on the strategy for selecting respondents, the researcher can choose a \textbf{P) probabilistic} (e.g., random selection) or \textbf{NP) non-probabilistic} (e.g., convenience, quota, snowballing) sample. This decision mainly affects the data analysis methods (as probabilistic samples are meant to be generalizable) and recruitment approaches (e.g., random selection) employed.
    \item[D2] \emph{How are the participants recruited?}\\On the one hand, \textbf{SS) self-selection} approaches allow for potential respondents to volunteer themselves which may introduce biases in the interpretation of the data.\\On the other hand, \textbf{PS) personalized selection} (such as invitation letters and more rarely authorization codes) require specific actions for the recruitment and management of the responses. This decision-making point is often interdependent of the kind of sample (D1).
    \item[D3] \emph{What type of survey instrument is designed?}\\\textbf{SA) Self-administrated} surveys are mainly distributed in the form of Web pages or printed questionnaires, thus the respondents fill out the data themselves.\\\textbf{IA) Interviewer-administrated} surveys include face to face or phone interviews where respondents provide the information to a researcher, who records the data. This decision not only drives the instrument design but can also heavily impact the execution stages (i.e., recruitment and response management).
    \item[D4] \emph{What response formats are collected?}\\Question structure types could be \textbf{OE) open-ended} and \textbf{CE) close-ended}. Open-ended questions are less restrictive allowing for respondents to use their own words, whereas close-ended questions are represented by scales that can be easily quantified. This decision determines the data analysis methods employed, i.e., qualitative or quantitative approaches. Often survey instruments include a mix of both question types, thus requiring both analysis approaches.
\end{enumerate}

\subsubsection*{C2. Recommended Practices}
\label{practices}

\begin{figure}
  \centering
  \includegraphics[width=1\textwidth]{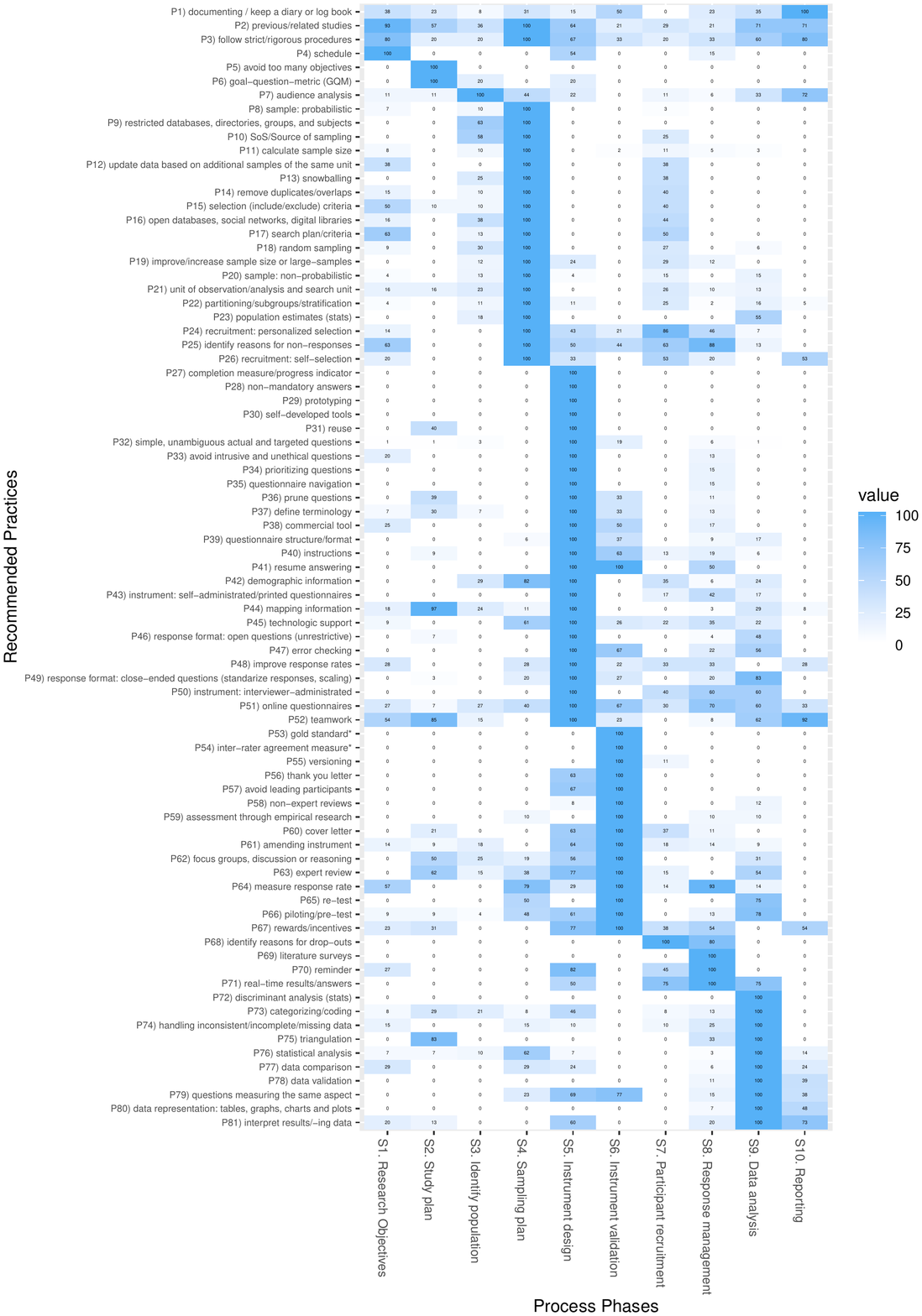}
  \caption{Co-occurrence matrix of recommended practices according to the process stages. Each cell contains a normalized coefficient, i.e., the highest co-occurrence value in each row is assigned a value 100 whereas the lowest value is 0. Cell shading illustrates the strength of the normalized co-occurrence coefficient, ranging from white (0) to blue (100).}
  \label{fig:coocurrence1}
\end{figure}

We identified a list of recommended practices for the survey research process (see Figure  \ref{fig:coocurrence1}). A stronger shade means that a higher number of citations were identified, however, the importance or the actual extent to which the practices and stages are related were not analyzed. 

Some practices are likely more relevant to a stage other than pointed by the cell shade. As an example, the recommended practice ``keep a diary/log book (P1)'' is strongly related to stage S10. Reporting, although this practice is often initiated in stage S1. Research Objectives.


The practices are mostly focused on specific stages of the survey process. However, they can also influence preceding and succeeding stages. The effect on the preceding stages is usually in the form of planning the strategy to be carried out. The impact on succeeding stages usually entails follow-up actions and the consequences of a given decision (see Section \ref{sec:process}).

\begin{enumerate}
    \item \emph{Research objectives:} Survey-based research is motivated by a specific goal. Thus it is important to state the research questions that correspond to such goal. The two main recommendations related to this stage are P1) to limit the scope, as this could impact upon the survey's complexity, and P2) to apply the goal-question-metric (GQM) approach to define its objectives. Moreover, the questionnaire items and collected data should be mapped to the research questions (P44).

    \item \emph{Study plan:} The need for designing the survey research is set at the beginning of the process, often along with the research questions \cite{kasunic2005designing}. The main suggested practices for this stage are to: P3) investigate related work; P4) define a set of procedures to guide the process; P5) develop a schedule plan for the stakeholders; and P6) start a diary or log book. The study plan should then be iteratively revised during the process, and the updates recorded in the log book (P1). This information is specially required for the reporting stage, at the end of the process.

    \item \emph{Identify and characterize the population:} Audience analysis (P7) is often employed to identify and select the characteristics of the population addressed by the research. This task has a strong effect on the sampling stage, in which the sources of sampling (P10) should be defined. Surveys often target potential participants at open databases (P16), but could employ restricted databases (P10) as an alternative or complementary source of sampling. Restricted databases should be investigated prior to the sampling stage.

    \item \emph{Sampling plan:} It is often employed in order to sample the population representatively. A sample plan should contain the sources of sampling (P10), units of observation and search unit (P21). The type of sample (P8 and P20) should potentially lead the decision for the data analysis methods employed. Other essential aspects to be considered are the P11) size of the sample and P19) how to manage large samples.

    Additional practices for this stage include to P14) remove the redundant units; P15) apply criteria for selecting the units of observation; P17) plan the retrieval of search units; and P22) partition the population according to the chosen characteristics. Strategies for recruitment (e.g., P8, P18, and P20) are likely to impact the participant selection stage.

    \item \emph{Instrument design:} A questionnaire or similar instrument is designed to gather data from the sample representative of the target population. Depending on the choice of distribution, the instruments can be P43) self-administrated, e.g., online forms (P51), or P50) interview-administrated e.g. interview or phone survey. They can be P29) prototyped, P30) implemented from the sketch, or acquired through P38) commercial tools or P31) reuse.

    Several recommendations to design and present an instrument are provided in the literature, e.g., avoid P33) intrusive and unethical questions, and P57) to lead the respondents; provide P27) a progress indicator, P35) questionnaire navigation, P40) instructions of use, P41) option to resume answering, and mainly P32) ask simple, unambiguous, actual and targeted questions. Responses can assume P46) open-ended or P49) close-ended formats.

    \item \emph{Instrument validation:} After design, the ability of the instrument to measure what is intended should be assessed. The most frequently cited approaches for the assessment are P66) piloting, P65) retest, P62) focus groups, and P63) expert or P58) non-expert reviews. Additionally, user-related metrics (e.g., usability, readability, time to respond) can result in improvements to the instrument design (P61). Ancillary documents supporting the recruitment stage should also be reviewed, e.g., cover letter (P60) and thank you letter (P56), likely providing incentives to the respondents (P67).

    \item \emph{Participant recruitment:} The strategies to select potential participants are previously defined in the sampling plan stage, such as P24) invitations and authorization codes, P26) self-recruitment, and P13) snowballing. By adopting proper actions and technology support, researchers can even investigate the potential threats to the process related to drop-outs (P68).

    \item \emph{Response management:} After distribution of the instrument to the selected participants, it is important to observe the response rate (P64) in order to identify the reasons for non-responses (P25). To ensure that the expected number of responses is achieved, researchers are likely to send reminders (P70) or to provide rewards for participation (P67).

    \item \emph{Data analysis:} Prior to the synthesis, the collected data should be validated (P78) in order to handle incomplete and missing values (P74). Furthermore, qualitative (e.g., P73) or quantitative (e.g., P76) analysis methods can be employed according to the survey's sample and response format. The results should then be P80) presented, P81) interpreted and likely P77) compared to particular subsets of the population. An additional suggestion to ensure their reliability is P79) to have more than one item measuring the same variable.

    \item \emph{Reporting:} The main practice related to the reporting phase is to produce an output of the information contained in the process documentation (P6). Ideally, the documentation is to be updated during the survey process, including the data analysis and results' interpretation. Both the related work (P3) and the adopted guidelines (P4) are used as additional information sources for this stage. Finally, it is important to consider the report's intended audience (P7).
\end{enumerate}

The frequency with which the practices occur in the segments of the text does not denote its importance for the process. The reasons to adopt a particular practice over another depends on the researchers' conscious decision supported by the guidelines employed.

\subsubsection*{C3. Rationales and Outcomes}
\label{rationales}

We define a rationale as the motivation to choose a particular practice. They are often described as desirable process attributes (e.g., cost-effectiveness, generalizability) or outcomes of such actions (e.g., minimize or introduce bias). As an example, to achieve generalizability, a researcher should utilize probabilistic sampling (P8) and estimates of the population size (P23).

\begin{figure*}[!ht]
  \centering
  \includegraphics[width=1\textwidth]{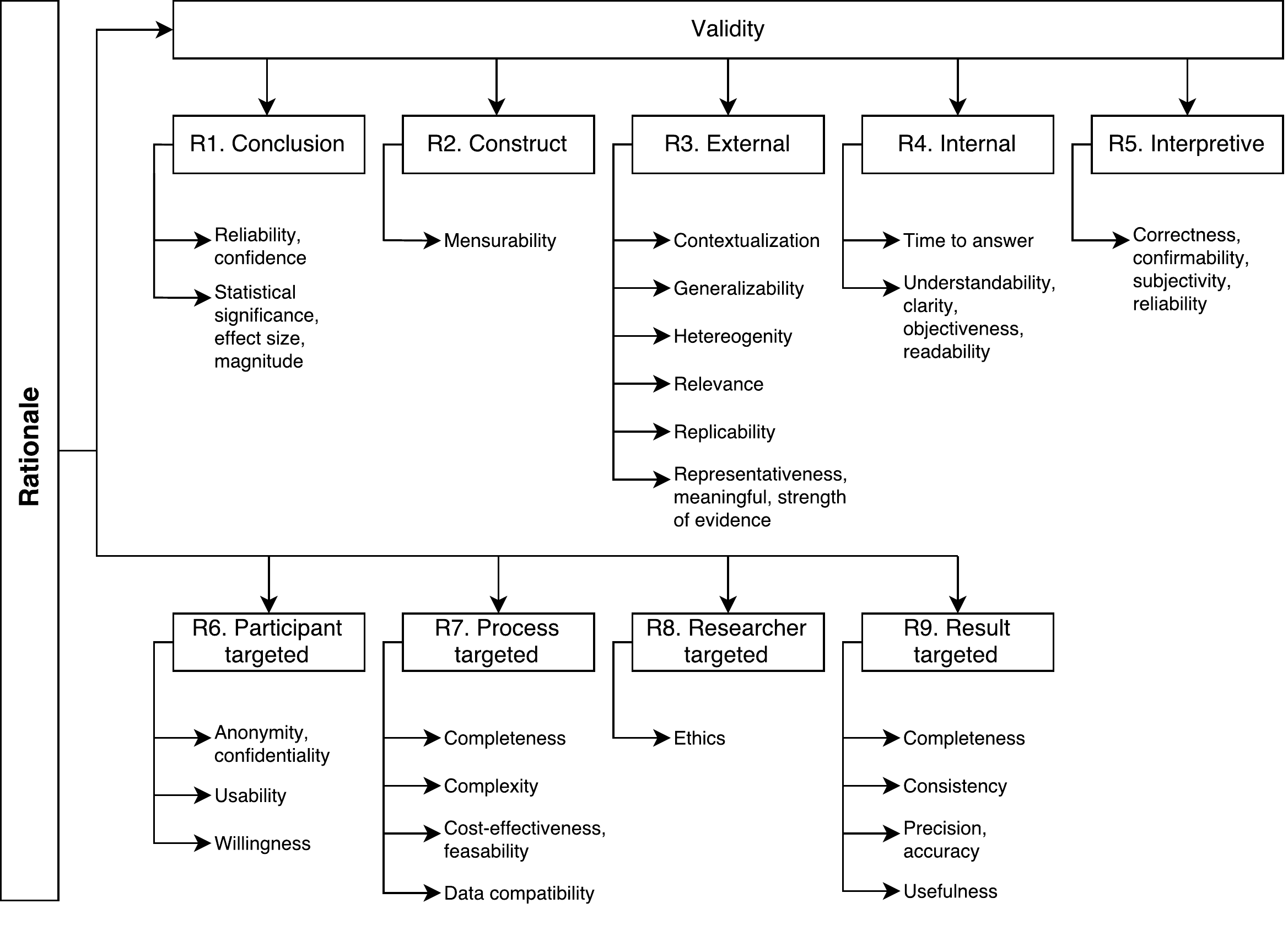}
  \caption{Rationale Aggregation Tree. Boxes represent the major categories and the related rationales are listed below. The topmost five categories are related to the validity of the research, whereas the bottom four are combined according to targeted aspects, i.e., participants, process, researcher and results.}
  \label{fig:rationalesTree}
\end{figure*}

Moreover, several of the rationales can be related to validity threats, i.e., concerns about the methodology in order to achieve valid conclusions \cite{petersen2013worldviews}. The additional rationales are not fully linked to validity, but can still positively or negatively affect it (e.g., willingness is likely to influence data quality, and thus validity). The coding resulted in a set of nine rationale categories, as shown in Figure \ref{fig:rationalesTree}.

\begin{enumerate}
    \item[R1] \emph{Conclusion validity:} the actual extent to which conclusions about the investigated relationship are true or correct. Survey-based research employing quantitative analysis methods are prone to significance, effect size, and magnitude factors. Moreover, the reliability or confidence of the results is inherent to the conclusion validity.
    \item[R2] \emph{Construct validity:} refers to the interaction between the underlying theory and measurement constructs, i.e., if the variables are actually measuring what they mean to. The main rationale in this category is mensurability, mainly addressed by the instrument validation stage.
    \item[R3] \emph{External validity:} the degree to which the results of the survey can be applicable to other scenarios, such as different contexts and strata of the population. Surveys are largely impacted by external validity factors, such as generalizability, replicability, and relevance to practice. Some major factors in this category are whether the sample is representative and heterogeneous to the overall target population.
    \item[R4] \emph{Internal validity:} represents an estimate of the degree to which conclusions about the investigated relationships can be drawn based on the measures and the research process. In survey-based research, the rationales in this category are mostly related to the sampling and instrumentation stages, e.g., understandability and time to respond.
    \item[R5] \emph{Interpretive validity:} related to the inference of the participants' opinions from the collected responses. Unlike the conclusion validity, the interpretive is more focused on the analysis of the qualitative data. Thus, factors such as correctness, confirmability, and subjectiveness play an important role in interpreting the data.
    \item[R6] \emph{Participant-targeted:} additional factors related to the respondents include concerns about anonymity and confidentiality, usability and willingness. Those factors are likely to impact the data quality, as they can positively or negatively influence the participants while answering the survey.
    \item[R7] \emph{Process-targeted:} the improvement of the process itself is a target of several rationales. Researchers carrying out surveys should pay special attention to cost-effectiveness, as their decisions are likely to require extra resources. Other practical considerations include the complexity of the instruments and techniques, completeness of the sampling sources, and compatibility of produced data.
    \item[R8] \emph{Researcher-targeted:} by providing their opinions, survey respondents trust that the gathered data will be processed responsibly. Thus, it is essential that the researchers are aware of potential ethical issues and their responsibilities regarding the survey process.
    \item[R9] \emph{Result-targeted:} one would expect a properly conducted survey process to produce useful results. Data validation tasks are meant to assess data consistency and completeness. Moreover, the precision of the results can be achieved by properly addressing a representative sample of the target population.
\end{enumerate}

Rationales are potential standards for quality of survey research. Moreover, by relating them to the practices can potentially support design decisions, e.g., the open-ended question format has implications on the interpretive validity from a researcher's perspective (R5); whilst standardized close-ended are leaning to conclusion validity (R1). Therefore, it is important to reflect upon the importance of different rationales while conducting the research process.

\subsubsection*{C4. Checklist Instrument}
\label{checklist}

The checklist had originally 38 items, distributed according to the survey process stages (see pre-evaluation checklist in Appendix \ref{sec:appendixChecklistv1}). At the end of each question, we also identify the related practices and rationales, which are intended to support reviewers while using the checklist. The rightmost column identifies the key decision points and related conditions that could impact the research (see Section \ref{sec:process}, e.g., D1:P means kind of sample: probabilistic as D1:NP is related to the non-probabilistic alternative).

Later, we complemented the reporting section of the checklist with more generic questions  selected from the checklist proposed by Dyb{\aa} and Dings{\o}yr \cite{dybaa2008empirical}. Those questions are meant to assess the quality of the evidence produced by empirical studies, regardless of the research method employed.

One can notice that several practices are presented within a stage not in accordance to the co-occurrence table (Section \ref{practices}). Although some of the practices require early planning, they could eventually be carried out in a later stage. Therefore, the checklist is organized in accordance with the stages in which those actions are more likely to take place.

Several process stages and their recommended practices are subject to decision-making (e.g., the kind of sample, instrument type). The choice should be guided by the motivations (i.e., rationales) and desired outcomes of the process. Due to this decision-making aspect, not all checklist items can be achieved to the same degree at once. We, therefore, rely on the researchers to prioritize the checklist items and hence made trade-offs according to their research goals.

\section{Step 2. Evaluation of the checklist}
\label{sec:evaluationProfessional}

In the second step of our work, we conducted a evaluation of the checklist in a research practice context, i.e., with researchers that published survey research papers. This evaluation intended to assess the appropriateness of the checklist, i.e., how well it addresses the needs of the research community to assess survey-based research. In particular, we were interested in evaluating the completeness, relevance and fairness of the checklist. To address these goals, we formulated three evaluation questions:

\begin{enumerate}
    \item[EQ1] \textbf{Completeness:} Is the checklist missing any important aspect?
    \item[EQ2] \textbf{Relevance:} Does the checklist contains items that are not relevant for SE research? and
    \item[EQ3] \textbf{Fairness:} Is the assessment using the checklist too lenient or stringent? 
\end{enumerate}

\subsection{Method}

The goal of our evaluation was to verify whether researchers agreed with our independent assessment of their work, and furthermore whether they though the checklist was complete and fair to assess survey research reports. To address such goal, we employed a mixed quantitative-qualitative approach, using our assessment through the checklist as the subject of study. Similar strategies for evaluating methods and tools in the SE context are described in \cite{kitchenham1997desmet}.

A set of conditions favor the decision to adopt such a evaluation approach within the research community, such as:

\begin{itemize}
    \item The subject of study is widely available, i.e., survey-based articles published in ESE journals and conferences. We can directly measure the compliance to the survey practices by assessing these objects with the support of our checklist.
    \item Researchers experienced with conducting surveys (i.e., corresponding authors of the above-mentioned papers) are likely to be interested in the checklist's potential use. Their expert judgment is also essential to identifying limitations and to gather improvement suggestions.
    \item Some of the benefits from using the checklist to assess research (i.e., rigour and fairness) are difficult to quantify. Thus, open-ended questions are more likely to provide a deep understanding of the opinions and reasoning of the experts regarding such aspects.
\end{itemize}

Our research evaluation process is based on a series of steps derived from a literature search \cite{kitchenham2015evidencebased} and survey-based recruitment and data collection \cite{kasunic2005designing}. Later, we analyzed the data according to both quantitative and qualitative synthesis procedures \cite{cruzes2011recommended}.

\subsubsection{Selection and recruitment}

\textbf{Search strategy.} At first, we identified survey-based articles that can be assessed using our checklist. We searched for potential candidates in nine venues (four journals and five conferences) publishing empirical research studies in SE, namely:

\begin{description}
    \item[TSE] Transactions on Software Engineering 
    \item[IST] Information and Sofware Technology 
    \item[ESEJ] Empirical Software Engineering Journal 
    \item[JSERD] Journal of Software Engineering Research and Development 
    \item[ICSE] International Conference on Software Engineering 
    \item[SEAA] Euromicro Conference on Software Engineering and Advanced Applications 
    \item[IWSM-Mensura] International Workshop on Software Measurement 
    \item[EASE] International Conference on Evaluation and Assessment in Software Engineering 
	\item[ESEM] International Symposium on Empirical Software Engineering and Measurement 
\end{description}

Despite well-established guidelines \cite{kitchenham2001principles,kasunic2005designing}, our checklist also incorporate practices mentioned in recent guidelines (e.g. \cite{linaker2015guidelines,de2015characterizing,de2015characterizing}). Thus, we opted for candidate papers that were published in the 5 most recent years (i.e., from 2012 to 2017), as they are more likely to incorporate such practices. From this database, we identified 3429 potential publications matching these characteristics.

\textbf{Selection process.} We further filtered the papers that mentioned the term ``survey'' in the title or abstract, thus narrowing the original list down to 177 candidates. We gathered these papers and selected them according to an inclusion criterion: \textit{Does the paper clearly reports survey-based research?} This resulted in 62 included papers.

\textbf{Recruitment.} Later, we invited by e-mail the corresponding authors of the selected papers to participate in our evaluation. Two of the corresponding authors have more than one paper in our candidate list, thus we sent 60 invitations related to 62 resulting papers. The invitation letter presented the goal and the context of the research and also described the assessment procedure (see evaluation procedure, below).

\textbf{Responses.} Three invitation e-mails could not be received with the given e-mail address. Out of the 57 authors who received an e-mail, 22 agreed to participate. One of them consented in assessing two of the papers we asked for and also provided an extra paper which was not part of our dataset. The additional paper was selected and aggregated to our list.

\subsubsection{Evaluation process}
\label{sec:evalProcess}


The process to evaluate our checklist consisted of:
\begin{enumerate}
    \item collecting the referred paper and applying the checklist to assessing it; 
    \item providing the corresponding authors with the filled out checklist so that they could review our assessment; and finally
    \item asking the corresponding authors to provide feedback regarding the checklist instrument and the resulting assessment.
\end{enumerate}

The resulting scores from our assessment using the checklist (see item 1, above) were aggregated into a dataset. This dataset was further analyzed in order to explore how many of the papers addressed each checklist item. Identifying patterns such as checklist items poorly addressed by most of the papers is essential for the next steps of our study. After receiving the participant's feedback, we compared theirs review to the patterns we identified.

Out of the 22 corresponding authors who agreed to participate, 12 provided us with a feedback, which consisted of:
\begin{itemize}
    \item a \textbf{review of our assessment}, in which the corresponding author can point out disagreements with our assessment, and and refine the assessment scores; 
    \item \textbf{response to three opinion questions} regarding the checklist, as follows:
    \begin{enumerate}
        \item Do you consider the checklist complete? If not, what should be included?
        \item Is there anything you would like to remove, or do you think it is irrelevant?
        \item Do you think our assessment by means of this checklist is fair? That is, was our assessment of the paper too rigid or too lenient?
    \end{enumerate}
\end{itemize}

\subsubsection{Data analysis}

Our analysis considered the feedback provided by the participants in the form of (1) a review of our assessment, and (2) answers to a set of opinion questions. 

In order to analyse the \textbf{review of our assessment}, we gathered the notes and comments provided by the participants regarding each of the checklist items. These notes were used to assess the \textit{completeness and relevance} (RQ1 and RQ2, respectively) of our checklist. In particular, we look for suggestions to improve the checklist, whether by removing, adding, or rephrasing. We responded to each comment and highlighted any action we took to improve the checklist based on the participants' feedback (see Appendix \ref{sec:appendixRejoinder}).

Furthermore, we assessed the \textit{fairness of our assessment} (RQ3) by computing the inter-rater agreement between the scores in our assessment and the ones reviewed by the corresponding authors. The inter-rater agreement is expressed in accordance with Cohen's kappa coefficient.

We also aggregated the participants' \textbf{answers to three opinion questions} into a common list\footnote{Available at \url{https://goo.gl/XE7wQF}.}. These open-ended answers comprise the respondents own phrasing and reasoning regarding the three topics of our evaluation (i.e. \textit{completeness, relevance, and fairness}).

We read each of the answers and assigned a value in a scale of yes/no/partial, representing their agreement with the question. We used both information types (i.e., assigned value and open-ended text) to answer our evaluation questions. Ultimately, we compared the participants' opinions with the findings from the review of our evaluation in order to identify recurrent themes.

\subsection{Threats to validity}

The checklist results from a systematic process to elicit recommended practices from survey guidelines in SE research. This process is not biased-free, however. In order to assess how well the checklist addresses the needs of the research community, we ought to evaluate its completeness and relevance, and fairness of the produced assessment.

\textbf{Construct validity.} A major threat to validity concerns the ability to assess the constructs with qualitative questions. We asked the participants to provide their own opinions regarding the checklist and our assessment. In particular, one of the participants questioned whether completeness could be assessed based on opinions.

Conventionally, open questions are associated to subjective responses, which is likely to constrain the analysis of data. To reduce such effect, we assigned an agreement value (using a 3-point scale) to the participant's opinion. Besides the participants' opinions, we support our findings with the scores resulting from our assessment. These scores are used to identify the practices often not reported or addressed.

\textbf{Interpretative validity.} Another potential threat to validity relates to the interpretation of the findings. In particular, we formulated three open-ended questions to collect participants' opinions. The questions themselves are not bias-free, as they are formulated to extract a positive/negative response. As an example, ``Do you consider the checklist complete?'' received more positive than negative answers. To decrease this threat, the data analysis and interpretation of our evaluation study were conducted by the first author and discussed with the other co-authors.

\textbf{Reliability.} Our great involvement in constructing the checklist is likely to introduce personal biases on our assessment scores. We aimed to mitigate these by building a traceable chain of evidence. First, we assessed the selected papers and recorded notes to support the given scores. We later asked the corresponding authors to review our scores and notes, and to refine any disagreement they identified. We further computed the inter-rater agreement between ours and the participants' scores, resulting in a very strong agreement (k = 0.91, according to weighted Cohen’s Kappa \cite{cohen1968weighted}).


\textbf{External validity.} Our selection process aimed to identify a diverse set of survey-based articles, i.e. surveys in different areas and/or surveys of different quality. The sample of papers collected covers a wide range of SE topics, e.g., testing, modeling, and industry practice. These papers were peer-reviewed, so we assume they present a rigorous and sound description from the survey process. This assumption is supported by the results of our assessment, in which the selected papers comply with 65\% of the items in our checklist. Thus, our sample is not diverse with regard to the methodological quality of the papers. Besides that, the participation of experienced researchers supports the generalization of our findings by expertise.

\subsection{Results}

\subsubsection{Our assessment using the checklist}
\label{sec:resultAssess}

We applied our checklist to assess 24 papers reporting survey research. Each of the checklist items was ranked as fully addressed (F), partially addressed (P), not addressed (N), or not applicable (NA). A summary of our assessment is presented in Table \ref{tab:summaryAssess}.

\newpage
\begin{footnotesize}
\begin{longtable}{cp{.4\textwidth}ccccc}

\caption{Summary of the combined scores obtained by the papers in our sample. Each row represents a checklist item, and the relative amount of papers (out of 24) ranked as fully addressed (F), partially addressed (P), not addressed (N), or not applicable (NA). The last column computes a compliance score based on how many papers address the related item.}
\label{tab:summaryAssess}\\

\renewcommand{\arraystretch}{1.2}

    \textbf{\#} & & \textbf{N} & \textbf{P} & \textbf{F} & \textbf{NA} & \textbf{Compl.} \\
    \hline
    \endfirsthead
    \multicolumn{7}{c}%
    {\tablename\ \thetable\ -- \textit{Continued from previous page}} \\
    \hline
    \textbf{\#} & & \textbf{N} & \textbf{P} & \textbf{F} & \textbf{NA} & \textbf{Compl.} \\
    \hline
    \endhead
    \hline
    \multicolumn{7}{r}{\textit{Continued on next page}} \\
    \endfoot
    \hline
    \endlastfoot
    
    \hline
    \multicolumn{7}{l}{\textbf{1. Research Objectives}}\\ \hline
    1A & Are the research question(s)\ldots & 1 & 0 & 23 & 0 & 95.8\% \\
    1B & Is the research context defined?\ldots & 1 & 0 & 23 & 0 & 95.8\% \\
    1C & Are the needs for the survey\ldots & 1 & 0 & 23 & 0 & 95.8\% \\ \hline
    \multicolumn{7}{l}{\textbf{2. Study plan}}\\ \hline
    2A & Is the survey process supported by guidelines?\ldots & 13 & 2 & 9 & 0 & 41.7\% \\
    2B & Is there a reflection on the need to update the research plan?\ldots & 19 & 0 & 5 & 0 & 20.8\% \\
    2C & Are the roles and responsibilities\ldots & 20 & 2 & 2 & 0 & 12.5\% \\ \hline
    \multicolumn{7}{l}{\textbf{3. Identify population}}\\ \hline
    3A & Is the population characterized\ldots? & 13 & 0 & 11 & 0 & 45.8\% \\
    3B & Is the size of the population\ldots & 19 & 1 & 4 & 0 & 18.7\% \\ \hline
    \multicolumn{7}{l}{\textbf{4. Sampling plan}}\\ \hline
    4A & Is the kind of sample\ldots defined? & 8 & 5 & 11 & 0 & 56.2\% \\
    4B & Is the sample size calculated\ldots & 7 & 1 & 16 & 0 & 68.7\% \\
    4C & Are the sources of sampling\ldots & 1 & 2 & 21 & 0 & 91.7\% \\
    4D & Are the strategies and criteria to select units\ldots & 10 & 0 & 14 & 0 & 58.3\% \\ \hline
    \multicolumn{7}{l}{\textbf{5. Instrument design}}\\ \hline
    5A & Is the type of instrument\ldots defined? & 1 & 1 & 22 & 0 & 93.7\% \\
    5B & Is the instrument design process\ldots & 5 & 2 & 17 & 0 & 75\% \\
    5C & Are the demographic questions \ldots & 2 & 2 & 20 & 0 & 87.5\% \\
    5D & Does particular care is taken to make the questions understandable\ldots? & 10 & 2 & 12 & 0 & 54.2\% \\
    5E & Is the number and order of the questions taken in consideration? & 16 & 1 & 7 & 0 & 31.2\% \\
    5F & Is there a reflection on the type of responses\ldots for the questions? & 4 & 1 & 19 & 0 & 81.2\% \\
    5G & If employing close-ended questions, are the standardized response\ldots & 1 & 3 & 20 & 0 & 89.5\% \\
    5H. & Is there a reflection on the adoption of additional sources\ldots & 18 & 2 & 4 & 0 & 20.8\% \\ \hline
    \multicolumn{7}{l}{\textbf{6. Instrument validation}}\\ \hline
    6A. & Is the validation process of the survey instrument detailed?\ldots & 6 & 0 & 18 & 0 & 75.0\% \\
    6B. & Is the instrument measuring what is intended?\ldots & 6 & 5 & 13 & 0 & 64.6\% \\
    6C. & In case of an electronic or online questionnaire, is the usability \ldots & 21 & 2 & 1 & 0 & 8.3\% \\
    6D. & Are the results of the instrument validation discussed?\ldots & 10 & 1 & 12 & 1 & 54.3\% \\ \hline
    \multicolumn{7}{l}{\textbf{7. Participant recruitment}}\\ \hline
    7A. & Are the strategies to select participants\ldots & 0 & 1 & 23 & 0 & 97.9\% \\
    7B. & Are the ancillary documents\ldots & 13 & 4 & 7 & 0 & 37.5\% \\
    7C. & If rewards or incentives to respondents are provided\ldots & 0 & 0 & 2 & 22 & 100\% \\ \hline
    \multicolumn{7}{l}{\textbf{8. Response management}}\\ \hline
    8A. & Are the responses monitored?\ldots & 4 & 2 & 18 & 0 & 79.2\% \\
    8B. & Is there any action to be taken in case of non-responses\ldots? & 16 & 0 & 5 & 3 & 23.8\% \\ \hline
    \multicolumn{7}{l}{\textbf{9. Data analysis}}\\ \hline
    9A. & Is the data validated\ldots & 16 & 1 & 7 & 0 & 31.2\% \\
    9B. & Is the method for data analysis\ldots & 2 & 2 & 20 & 0 & 87.5\% \\
    9C. & If statistical analysis is employed, is the hypothesis testing process\ldots & 0 & 1 & 15 & 8 & 96.9\% \\
    9D. & If using qualitative synthesis\ldots & 0 & 0 & 10 & 14 & 100\% \\
    9E. & If a stratified sample is defined\ldots & 0 & 0 & 3 & 21 & 100\% \\ \hline
    \multicolumn{7}{l}{\textbf{10. Reporting}}\\ \hline
    10A. & Are the instrument and ancillary documents accessible\ldots & 5 & 1 & 18 & 0 & 77\% \\
    10B. & Has a discussion of both positive and negative findings\ldots & 0 & 1 & 23 & 0 & 97.9\% \\
    10C. & \ldots Are limitations of the study (e.g. threats to validity) discussed? & 0 & 3 & 21 & 0 & 93.7\% \\
    10D. & Are the conclusions justified\ldots & 0 & 1 & 23 & 0 & 97.9\% \\ \hline
    \hline
    & \textbf{Mean} & 11.2 & 2.04 & 21.88 & 2.88 & 65\% \\ 
    \hline
\end{longtable}
\end{footnotesize}

For each assessed item, we also added notes for possible improvements in the study's documentation, e.g., due to missing information. As an example, in relation to the checklist item 2A, which assess the detailed procedures when designing a survey, we provided the following note to one of the participants: ``\textit{The paper cited guidelines to survey research to characterize the sample and recruitment. It is not clear if the method provided in the guidelines are followed thought all the research process.}'' In order to preserve the anonymity, we do not report the complete notes here. They were however shared with the corresponding authors.

The last column of the table presents a compliance score, i.e., the relative amount of papers that addresses the related checklist item. A score of 100\% means that all papers were rated ``F''. The NA ratings are not computed, and each P counts as half of a full score. The compliance score does not take into consideration the importance of different items for the research. Thus it should be interpreted merely as an account of possible improvements to be taken into consideration.

\textbf{EQ1. Completeness:} Overall, our sample of papers complies with 65\% of the items in the checklist. Some of the checklist items and groups presented better compliance, such as the items related to the research objectives (1A to 1C) and three out of four items related to reporting (10B to 10D). One expects any research work, regardless of research method employed, to meet these requirements.

\textbf{EQ2. Relevance:} Three checklist items are fully addressed by all the papers assessed. They cover practices such as incentives to responses (7C), qualitative synthesis (9D) and stratified data analysis (9E). These items are optional, and thus the assessment is rated not applicable (NA) for all the papers that do not employ such strategies. These results imply that researchers applying such strategies are likely to report them explicitly.

Among the checklist groups that are more scarcely addressed are:
\begin{enumerate}
    \item[2)] study plan; 
    \item[3)] identify the population;
    \item[6)] instrument validation; and
    \item[8)] response management.
\end{enumerate} 

The low compliance scores show that the same kind of information is missing in several assessed studies. This implies that some of the recommended practices proposed by the guidelines are not followed. If we consider that this sample of papers is a good representation of the overall survey-based research in SE, the low-compliance items point out to gaps that should be part of wider discussion so to see if they are relevant in survey research.

\textbf{EQ3. Fairness:} We later compared the participants' scores to ours via inter-rater agreement. The resulting weighted Cohen’s Kappa coefficient k = 0.91 \cite{cohen1968weighted}, suggesting a very strong level of agreement. We assume that two reviewers using the checklist independently are not likely to achieve such stronger agreement. However, the results showed that the corresponding authors judged the assessment as mostly fair. This is reinforced by the authors' answers the opinion questions (see Section \ref{sec:resultOpinions}).

\subsubsection{Evaluation by the corresponding authors}
\label{sec:resultOpinions}

After assessing all the papers, we provided the corresponding authors with the filled out checklist and our notes. We then asked them to provide feedback based on our assessment. Out of the 22 corresponding authors contacted, 12 replied to our request, providing feedback regarding the completeness of the checklist, irrelevant checklist items, and fairness of our assessment.

We addressed the participants' comments individually, responding to each issue in need of due attention and detailing the actions we took to improve the checklist. A subset of our responses are provided in Appendix \ref{sec:appendixRejoinder}, and the complete set is available online in \url{https://goo.gl/YDj1XA}.

\textbf{EQ1. Completeness:} Most of the participants (7 out of 12) agreed that the checklist was complete and included all the main aspects of survey-based research. Two participants thought that the checklist was partially complete, and it could be improved by clarifying a few items.

One participant highlighted their confidence that our method of creating the checklist was grounded in methodological publications, such as \cite{kitchenham2001principles}. This information was not provided beforehand, so we assumed that the participant is familiar with such work, thus relating our checklist items to the recommended practice described in Kitchenham's guidelines \cite{kitchenham2001principles}.

The three remaining participants who did not agree with the checklist completeness, raised issues such as:
\begin{itemize}
    \item internal and external validity are not completely addressed in relation to the sampling plan and the instrument validation;
    \item more details are needed for novice researchers using the checklist; and
    \item validating completeness is not possible as an opinion.
\end{itemize} These aspects are addressed individually in our feedback document (see Appendix \ref{sec:appendixRejoinder}), as mentioned above.

\textbf{EQ2. Relevance:} Three participants mentioned irrelevant checklist items they believed should be removed: 
\begin{enumerate}
    \item[2C)] the checklist item addressing research roles and responsibilities was considered irrelevant for the report, but it could be part of the research plan (2B);
    \item[6A/6C)] these two items should be combined, as they both address the instrument validation;
    \item[5H)] using additional sources for data collection is optional, therefore if not mentioned in the paper it should be rated NA; and
    \item[7B)] to provide ancillary documents (e.g., cover letter, invitation letter) is irrelevant to the research report.
\end{enumerate}

The only issue raised by more than one participant is related to unifying 6A and 6C. The results of our assessment point out that most of our sample studies are in compliance with 6A (75\%), but just a few (8.3\%) actually address item 6C. We think that it is important to keep these two aspects separated, thus making explicit the needs for validating the usability of the questionnaire (see e.g., recommended practices P27, P28, P35, P39, P40, P41, P45). All the issues abovementioned are discussed in our feedback document (see Appendix \ref{sec:appendixRejoinder}).

\textbf{EQ3. Fairness:} Most participants (9 out of 12) considered our assessment fair. Two of them also mentioned that despite rigid, the assessment was fair. Another one highlighted the need for instruments that promote rigorous assessment of the research methods. None of the participants described our assessment as completely unfair, although three of them pointed out that items we missed in our assessment were limitations to fairness.

We noted that two participants mentioned the lack of information due to size limitations of the publication. This issue is further highlighted in the comments of other participants (see Appendix \ref{sec:appendixRejoinder}). We sympathize with the participants' concern regarding a fair assessment due to the size limitation. However, we stress the importance to provide all the details needed to properly assess the research based on its report. As a recommended practice, researchers are encouraged to make additional information (e.g., research diary, questionnaire instrument, ancillary documents) accessible to the target audience.

\section{Discussion}
\label{sec:discussion}

\subsection{Checklist Usage}
\label{sec:hotouse}

In order to assess survey-based research, reviewers can employ the proposed checklist. Prior to assessment, we suggest verifying the availability of research process information (i.e., research report, survey instrument, and ancillary documents). Thereafter, each checklist item should be carefully read and then evaluated with respect to whether the question can be answered and was reflected on in the research report.

Several checklist items comprise two or more nested questions. Those items are intertwined and should not be assessed separately. Moreover, the checklist items can be addressed as partial coverage, due to the higher level of abstraction where answer is likely to be subjective. In such cases, we rely on the reviewers' best judgment regarding the adoption of partial scores (i.e., 0.5). 

It is possible to derive a scoring measure based on the checklist marks (e.g., 23 out of 38). However, we do not encourage the simple aggregation of scores in such a way, as it is likely to lead to a loss of assessment information. We suggest reviewers report the reasoning to score each question, thus highlighting the strengths and weaknesses of the assessed survey.

\subsection{Implications to Research}
\label{sec:implications}

The objective of our checklist is twofold: first, it is intended to audit reported survey-based research; and second, to support researchers in making research design decisions and reporting them. Ideally, both the researchers employing the checklist to plan and report their studies and the reviewers assessing the same research should obtain similar scores.

Alternatively, this reflexive checklist can be used to improve the survey process; researchers are encouraged to think and reflect upon the questions they are aiming to use. In particular, trade-offs have to be made. The completeness of the survey as well as the ability to obtain a large and representative sample are desired, but also costly. Thus, as highlighted in the survey guidelines, the research process decisions have to be reflected upon with respect to cost-effectiveness. This is not to say that researchers should aim at minimizing the cost, but rather reflect on what is needed to fulfil the research goals. 


Reviewers using our checklist are strongly encouraged to report the resulting scores along with their reflections about the checklist itself. We also foster independent evaluations to verify the appropriateness of the checklist to assess survey-based research by the research community.

Finally, the proposed checklist is intended to assess survey-based research in SE, but it has the potential to address different domains' studies (e.g., social sciences). It is important to identify the differences of the survey-based process employed in SE and in other fields, thus evaluating the checklist in a cross-domain study.

\subsubsection{Research Practice}

During the checklist evaluation, we assessed 24 papers reporting survey research. The results of our assessment (Section \ref{sec:resultAssess}) point to a list of recommended practices (see Section \ref{practices}) that are scarcely addressed. We believe that by communicating these insights to the community, we can encourage researchers to consider the recommended practices in their research. The scarcely addressed practices are:

\begin{enumerate}
    \item[2.] \textit{Study plan:} A research plan or log book (see recommend practice P6) is important to guide the research efforts. This protocol should detail the responsibilities of each stakeholder (P52) and a timetable (P5). The document should be updated as new information becomes known, and ultimately make it accessible by the end of the research.
    \item[3.] \textit{Identify the population:} Very often the demographics of the participants are described, but scarce information is provided regarding the target population. An audience analysis (P7) is likely to identify and supply these characteristics, and the census of practitioners and institutions could provide estimates of the population size (P23).
    \item[5.] \textit{Instrument design:} When designing the data collection instrument, one should carefully consider the order (P34) and amount (P36) of questions. Additional sources of data (P69), such as work repositories can provide means to cross-check the results.
    \item[6.] \textit{Instrument validation:} In general, the reports stated that some kind of validation of the questionnaire (e.g., a pilot) was performed. When employing online surveys, we should ideally check the usability of the questionnaire instrument. A series of practices (see e.g. P27, P28, P35, P39, P40, P41, and P45) can be used as a guideline or checklist to this validation. Furthermore, it is also important to report the improvements made as resulting of the instrument validation (P61).
    \item[7.] \textit{Participant recruitment:} Besides the need for making the research plan available, it is suggested to provide access to the standardized communication with participants. These documents include the invitation and cover letters (P60) as well as follow-ups and thank you letters (P56).
    \item[8.] \textit{Response management:} Besides the response rates, it is valuable to report any strategy used to improve responses, such as reminders (P70) or searching for additional databases (P16). These strategies are likely to affect the sample size, thus we should ideally discuss the implications of them for the study validity.
    \item[9.] \textit{Data analysis:} Before the analysis, researchers are encouraged to validate the data (P78) and check for inconsistent, incomplete or missing information (P74). There are several strategies to deal with missing data, from discarding to the imputation based on statistical models. Ideally, the implications due to employing such approaches should be described.
\end{enumerate}

\subsection{Comparison with Related Work}
\label{sec:comparison}

We previously identified related studies providing checklists for assessing empirical research in SE (see Section \ref{sec:checklists}). The existing checklists in SE target mostly experiments \cite{JedlitschkaP05,KitchenhamSBBDHPR10} and case study research \cite{WieringaCDMP12}. Our proposed checklist is intended to address the issues of survey-based research and its specific stages, e.g., sampling, instrument design, and recruitment.

Our resulting checklist can be comparable to Stavru's set of criteria to assess the thoroughness of surveys \cite{stavru2014critical}. Similar to their work, we also systematically derived our instrument from the literature. We detailed herein the empirical processes to construct our checklist.

Our checklist differs from Stavru's criteria, as it covers a larger set of practices and provide instructions based on the literature. The recommended practices often taken into account further activities of the survey research process. As an example, in relation to the research objectives, our checklist item A1: \textit{``Are the research objective expressed in measurable terms? E.g. as research questions, or using the goal-question-metric approach.''} In this particular case, the objectives should describe what is intended to be measured, thus favoring data analysis, as proposed by \cite{kitchenham2001principles,kasunic2005designing,linaker2015guidelines,ciolkowski2003practical}.

The items in our checklist are not weighted, as the relative importance of each practice depends on the survey process, and researchers are foster to reflect on the key decision points. Besides that, we evaluated our checklist with experienced researchers. The evaluation identifies room for improvement, and we provide an updated post-evaluation version in Appendix \ref{sec:appendixChecklist}.

More generic checklists \cite{Wieringa12,dybaa2008empirical} share some similarities with our proposed checklist. Similar questions are mostly related to the research objectives and reporting stages. This is expected since some actions (e.g., formulate research questions, define the scope, discuss the results and limitations of research) are inherent to all empirical research.

We considered the question formulation of the generic checklists when phrasing our proposed instrument. Researchers who used those similar checklists are likely to recognize the overlaps. This could be beneficial, as they can employ identical reasoning when assessing the common items. Similar questions provide the opportunity to compare the checklists or the studies assessed through them.

\section{Conclusions}
\label{sec:conclusions}

In this paper, we described a process to derive an instrument to assess survey-based research and its further evaluation. The motivation for such is grounded in the increased usage of survey-based research in SE. In a previous study, we identified several guidelines supporting the survey processes, but no instrument provided a checklist to assess their quality.

A set of 14 methodological papers provided qualitative data that was collected and analyzed through thematic analysis. The resulting themes resulted in sets of practices and rationales for carrying them out. We built the proposed checklist based on those extracted themes, reporting it through 38 questions organized by the survey process stages.

Later, we employed an empirical evaluation approach to collect experts' opinions of our checklist. We provided the experts with an assessment produced by applying the checklist to their own reported surveys. Overall, our instrument was evaluated as complete and the assessment as rigorous and fair. Issues regarding understandability and subjectivity of the checklist items were collected and based on this feedback, we update our proposed checklist.

We believe that the empirical software engineering community can benefit from our checklist for survey research. It can be a valuable asset for both researchers conducting and reporting survey-based studies, and for reviewers auditing survey reports.

As future work, we plan to investigate the potential benefits of using of the checklist by independent reviewers. We also intend to compare our checklist with other assessment tools (e.g. \cite{JedlitschkaP05,KitchenhamSBBDHPR10,WieringaCDMP12}) with respect to quality standards for empirical research.



\bibliography{references}

\bibliographystyle{abbrv}

\newpage
\begin{appendix}
    \section{Appendices}
    
    \subsection{Survey assessment checklist (pre-evaluation)}
    \label{sec:appendixChecklistv1}
    
    \begin{figure*}[!ht]
      \centering
      \includegraphics[width=.95\textwidth]{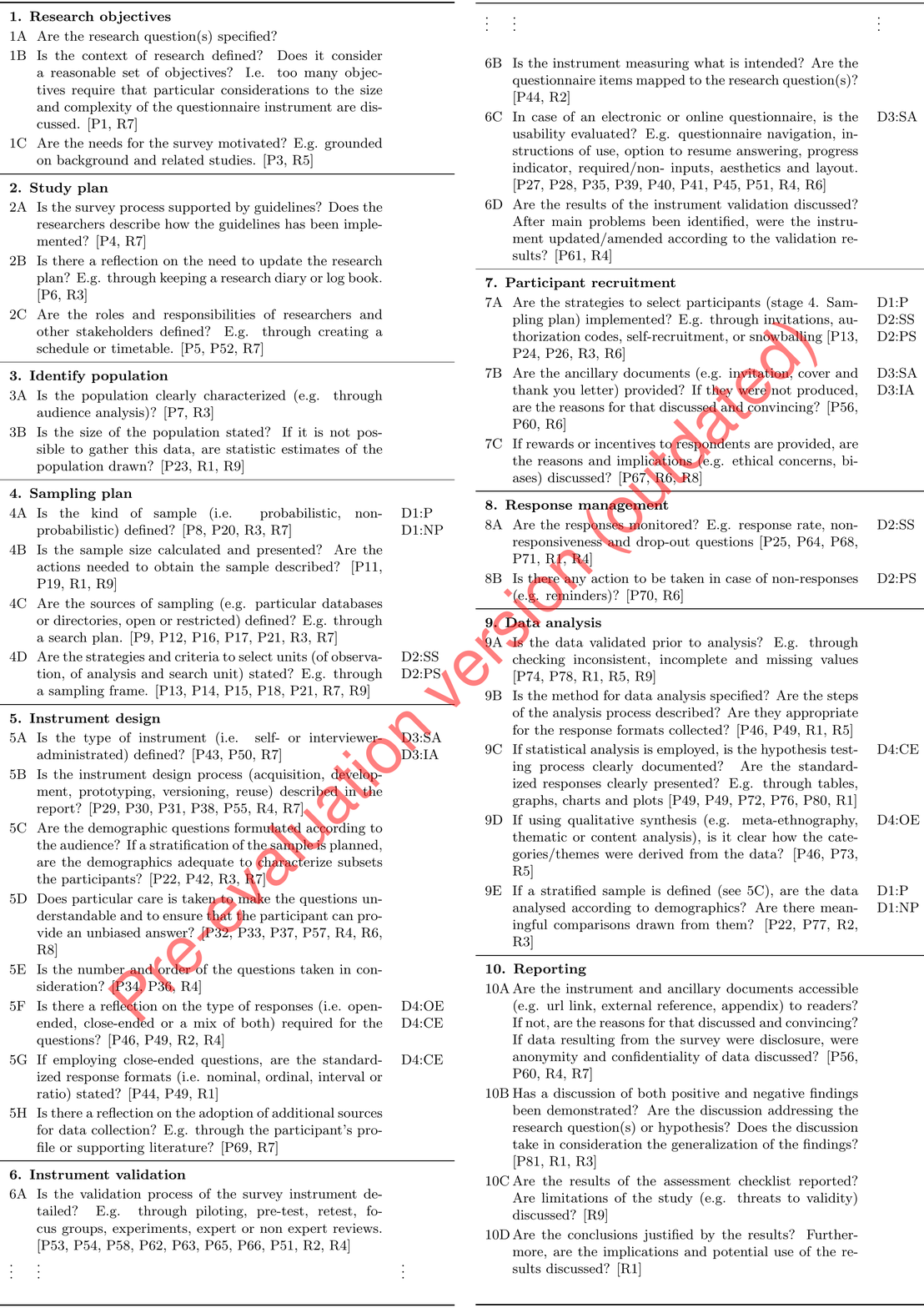}
      \caption{Survey assessment checklist proposed. This pre-evaluated version is later improved and updated (see Appendix \ref{sec:appendixChecklist}).}
      \label{fig:instrumentv1}
    \end{figure*}
    
    \newpage
    \subsection{Survey assessment checklist (post-evaluation)}
    \label{sec:appendixChecklist}
    
    \begin{figure*}[!ht]
      \centering
      \includegraphics[width=1\textwidth]{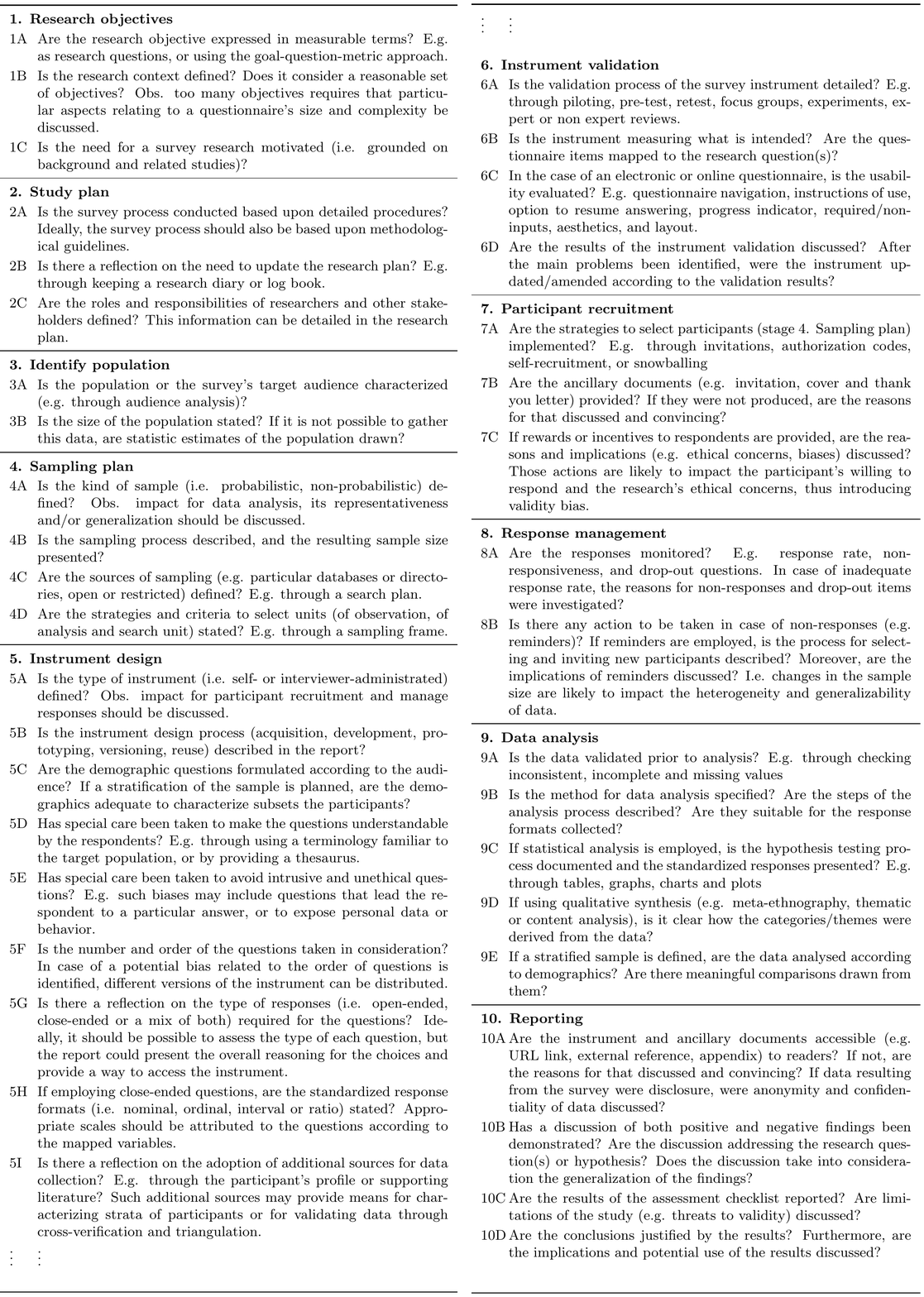}
      \caption{Survey assessment checklist after evaluation (see Section \ref{sec:evaluationProfessional}). A digital version of the checklist is available at \url{https://tinyurl.com/se-survey-checklist}.}
      \label{fig:instrumentv2}
    \end{figure*}
    
    \subsection{Suggestions to improve the checklist (excerpt)}
    \label{sec:appendixRejoinder}
    
    Here we present a sample of the feedback provided by the participants (i.e., corresponding authors) of our evaluation in the professional context (Section \ref{sec:evaluationProfessional}). The comments are listed according to the checklist item they are related to. For each comment, we present our responses and actions we took to address the mentioned topics. A complete list detailing all the comments is provided at \url{https://goo.gl/jNXx7U}.
    
    \begin{enumerate}

        \item[1B)] Two comments regarding the understandability of this checklist item: 
        \begin{enumerate}
            \item \textit{What type of context and limitations should be described? Many questions would benefit from having a more detailed guide along with the checklist.}; and 
            \item \textit{The term "limitations of scope" may be misleading. Reading quickly I first thought that you referred to whether the study scope has some limitations to be able to answer RQs (related to study validity).}
        \end{enumerate}
        \textbf{Response:} We agree that term “limitations of scope” can leave room for interpretations. It could also be misleading, as limitations are often described as threats to the validity of a study.\\
        \textbf{Action:} To improve the checklist understanding, we rephrased item 1B, as follows: ``Is the research context defined? Does it consider a reasonable set of research objectives? Obs. too many objectives requires that particular aspects relating to a instrument’s questionnaire’s size and complexity be discussed.''.
        
        \item[2)] Two comments regarding the description of this checklist item: 
        \begin{enumerate}
            \item \textit{The sub-questions for me do not address the main question of whether a survey is appropriate. 2A-C are more about what is reported, rather than whether survey is the right method. That for me is more about whether other approaches were considered etc. Roles and responsibilities I would generally not note in a paper.}; and 
            \item \textit{(...) I am not convinced that the three questions that are included in this category would be enough to assess if a survey study research design is appropriate to address its research aims (as it is stated in the question). The fact that guidelines are followed, a research diary is kept and responsibilities are defined does not guarantee that the research design is appropriate to answer the research questions. (...)}
        \end{enumerate}
        \textbf{Response:} We agree with the authors that the description of checklist item 2 does not match what is assessed in sub-items 2A-2C. These sub-items assess whether the study plan is accessible and complete, instead of ``appropriate''.\\
        In order to assess whether the survey design is appropriate to address its research objectives, we designed a specific question (see 6B), that assess if the questionnaire items are related to the research questions described in the study plan.\\
        The recommended practices for the checklist item 2 are: (i) to provide a survey plan document; (ii) to report the guidelines used; and (iii) to detail the responsibilities of each researcher. These aspects are important to allow for the study to be reviewed and replicated.\\
        \textbf{Action:} We updated phase 2’s description to match what is assessed by items 2A-2C: ``Study plan: Is the survey design accessible and complete?''.
        
        \item[2C)] Two comments regarding the relevance of this checklist item: 
        \begin{enumerate}
            \item \textit{Is this information really relevant for the report? I think this information would fits better into a protocol or research plan than in the report itself.}; and 
            \item \textit{It sounds irrelevant (e.g., how relevant it is for assessing the survey itself to know timetables)?}
        \end{enumerate}
        \textbf{Response:} We agree with the authors that a schedule or timetable is not relevant for assessing the quality of the survey. However, these artifacts can provide means to assess the roles and responsibilities of researchers in the survey process \cite{kasunic2005designing}. As suggested by one of the participants, the information regarding the roles and responsibilities of each researcher could be provided in a survey plan document. We highlight here the need to make this document accessible to reviewers \cite{cater2005addressing}.\\
        \textbf{Action:} We removed the references to schedule and timetables in the item 2C, instead stating of that ``This information can be detailed in the research plan (see item 2B)''.
        
        \item[5H)] Three comments regarding the understandability of this checklist item: 
        \begin{enumerate}
            \item \textit{I didn't understand it.}; 
            \item \textit{It makes no sense for me. If man don’t mention another source of information it means there is not. Why do you assume that there is another unmentioned source?}; and 
            \item \textit{I don't fully follow this question. Do you mean data triangulation so findings from the survey are triangulated with other data?}
        \end{enumerate}
        \textbf{Response:} The authors pointed out a very important understanding issue. Our proposed checklist accounts for the need to discuss if additional sources are required, e.g., to characterize stratas of the participants, or to cross validate the data related to the investigated phenomenon from multiple sources \cite{kasunic2005designing,linaker2015guidelines}. As an example, after the survey, the findings can be compared to other sources, such as personal profile information or related work.\\
        \textbf{Action:} We added a note on 5H making explicit the reasons to adopt additional sources of data collection, as follows: ``Such additional sources may provide means for characterizing strata of participants or for validating data through cross-verification and triangulation''.

        \item[10A)] \textit{Is data disclosure / open data also a criterion? I think it should be as people should be pushed in the general direction of open science to foster reproducibility.}\\
        \textbf{Response:} The author suggests that, besides the ancillary documents we already mentioned in the checklist, the resulting data from the survey is also make available. We see the value on that, but acknowledge potential implications due to anonymity and confidentiality that should be taken into consideration. Therefore, we rely on the judgment of researchers conducting the survey to provide a discussion on such aspect.\\
        \textbf{Action:} We added a note on 10A making explicit that the data disclosure can be provided and thus should be discussed. The new item is ``If data resulting from the survey are disclosure, does a consideration about the anonymity and confidentiality of data is discussed?''.

        \item[10A)] Two additional comments regarding applicability of this checklist item: 
        \begin{enumerate}
            \item \textit{I can't imagine including things like invitations, thank you notes, etc. My ethics approval process requires so much documentation, I would need a separate 10 pages just for all of the information I provide to participants. }; and 
            \item \textit{not all materials can be added to an appendix especially with paper length limitation.}
        \end{enumerate}
        \textbf{Response:} We agree with the author that limitations due the publication size are likely to constrain the amount of information provided in a paper. As an alternative, the additional documents mentioned in 10A could compose a “survey research package”, available as a web reference provided in the paper \cite{cater2005addressing}.

    \end{enumerate}
    
\end{appendix}

\end{document}